\documentclass[prl,aps,reprint,nobibnotes,twocolumn]{revtex4-1}

\usepackage{graphicx}%
\usepackage[version=4]{mhchem}
\usepackage{seqsplit}

\newcommand{\grant}[1]{{\seqsplit{#1}}}

\begin{document}

\title{Generation of Spatiotemporal Vortex Pulses by Simple Diffractive Grating}

\author{Zhiyuan Che,$^{1}$}
\author{Wenzhe Liu,$^{2}$}
\email{e-mail: wliubh@connect.ust.hk (W. L.); lshi@fudan.edu.cn (L. S.); phchan@ust.hk (C. T. C.);  jzi@fudan.edu.cn (J. Z.)}
\author{Lei Shi,$^{1,3,4}$}
\email{e-mail: wliubh@connect.ust.hk (W. L.); lshi@fudan.edu.cn (L. S.); phchan@ust.hk (C. T. C.);  jzi@fudan.edu.cn (J. Z.)}
\author{C. T. Chan,$^{2}$}
\email{e-mail: wliubh@connect.ust.hk (W. L.); lshi@fudan.edu.cn (L. S.); phchan@ust.hk (C. T. C.);  jzi@fudan.edu.cn (J. Z.)}
\author{Jian Zi$^{1,3,4}$}
\email{e-mail: wliubh@connect.ust.hk (W. L.); lshi@fudan.edu.cn (L. S.); phchan@ust.hk (C. T. C.);  jzi@fudan.edu.cn (J. Z.)}

\affiliation{$^{1}$State Key Laboratory of Surface Physics, Key Laboratory of Micro- and Nano-Photonic Structures (Ministry of Education), and Department of Physics, Fudan University, Yangpu District, Shanghai, 200433, China}
\affiliation{$^{2}$Department of Physics, The Hong Kong University of Science and Technology, Clear Water Bay, Kowloon, Hong Kong, 999077, China}
\affiliation{$^{3}$Institute for Nanoelectronic Devices and Quantum Computing, Fudan University, Yangpu District, Shanghai, 200438, China}
\affiliation{$^{4}$Collaborative Innovation Center of Advanced Microstructures, Nanjing University, Gulou District, Nanjing, 210093, China}

\begin{abstract}
  Spatiotemporal vortex pulses are wave packets that carry transverse orbital angular momentum, exhibiting exotic structured wavefronts that can twist through space and time. Existing methods to generate these pulses require complex setups like spatial light modulators or computer-optimized structures. Here, we demonstrate a new approach to generate spatiotemporal vortex pulses using just a simple diffractive grating. The key is constructing a phase vortex in frequency-momentum space by leveraging symmetry, resonance, and diffraction. Our approach is applicable to any wave system. We use a liquid surface wave platform to directly demonstrate and observe the real-time generation and evolution of spatiotemporal vortex pulses. This straightforward technique provides opportunities to explore pulse dynamics and potential applications across different disciplines.
\end{abstract}

\maketitle

\section*{Introduction and theory}

\par Spatiotemporal vortex pulses (STVPs) are structured wave packets that possess intrinsic optical angular momentum transverse to their propagation direction \cite{dror2011symmetric, bliokh2012spatiotemporal, jhajj2016spatiotemporal, chong2020generation, bliokh2021spatiotemporal, hancock2021mode, cao2022non, bliokh2023orbital, ge2023spatiotemporal}, imparting a twisting wavefront that spins like a wheel through space and time. These exotic pulses are the spatiotemporal analogs of widely studied vortex beams \cite{allen1992orbital, bliokh2006geometrical, dennis2009singular, chen2016geometric, huang2020ultrafast, yang2021generation, ni2021multidimensional, zhang2022multiplexed, wang2023coloured}. Unlike vortex beams, confined to three dimensions, STVPs can exist in two dimensions, like in surface waves. This unique property expands their potential applications beyond those possible with vortex beams alone, including spatiotemporal differentiating \cite{doskolovich2022spatiotemporal, zhou2023electromagnetic}. However, existing STVP generation techniques typically rely on complex wavefront-shaping devices like spatial light modulators \cite{chong2020generation, ge2023spatiotemporal} or specially engineered metasurfaces requiring extensive optimization \cite{wang2021engineering, zhang2023observation, zhou2023electromagnetic}.

\par Planar periodic structures, e.g. photonic crystal slabs, are ideal candidates for spatiotemporally modulating these wheel-like wave packets because they offer frequency ($\omega$) and in-plane wavevector ($\mathbf{k}_\parallel$) degrees of freedom \cite{johnson1999guided, fan2002analysis, fan2003temporal, guo2020squeeze, wang2020generating, wang2021shifting}, the reciprocal spaces of time and position. Modulation in the  $\omega$-$\mathbf{k}_\parallel$ domain directly manifests in the pulse's space-time properties \cite{wang2021engineering, doskolovich2022spatiotemporal, zhang2023observation, zhou2023electromagnetic}. Introducing a phase vortex in the $\omega$-$\mathbf{k}_\parallel$ domain generates an STVP. Here, we propose \emph{a general approach} to generate STVPs using just a simple diffractive grating. The key insight is harnessing symmetry, resonance, and diffraction of waves to construct a phase vortex in frequency-momentum space that directly manifests as a twisting pulse in time and space. We experimentally demonstrate this approach and visualize generated spatiotemporal vortex pulses in real-time using a surface wave platform \cite{dingemans1997water, hu2005refraction, zhao2021fast, han2022water, qin2023superscattering, zeng2023achieving}. Furthermore, we show that by incorporating additional resonances, our approach can generate multiple interacting singularities within a single STVP.

\begin{figure}[ht]
  \centering
  \includegraphics[width=0.5\textwidth]{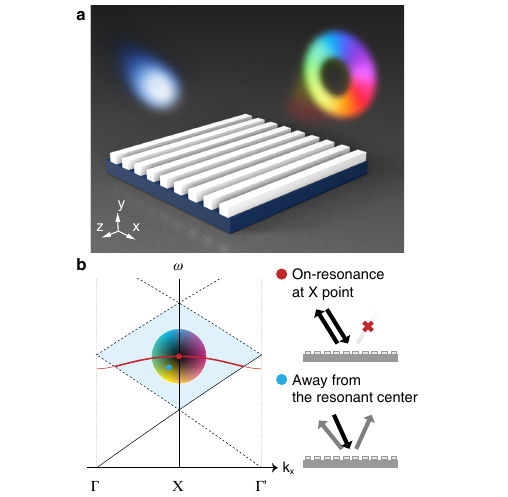}
  \caption{Diffractive generation approach of spatiotemporal vortex pulses (STVPs) by mirror-based grating. (a) A STVP centered at a specific frequency can be generated from gaussian pulses by a simple mirror-based resonant grating. (b), Diffraction and the mirror symmetry of the grating gives rise to this wave phenomenon. The blue-shaded region corresponds to the momentum-frequency range where the grating supports only the specular reflection and the -1st-order diffraction (i.e., two scattering channels). A phase singularity in specular reflection is formed at the X point in momentum space due to resonance-induced total-retroreflection. }
  \label{fig:1}
\end{figure}

\par Our approach has three simple prerequisites for the structures [Fig. \ref{fig:1}(a)]. First, periodicity in one or more dimensions that supports resonant band structures. These resonances play a key role in producing the $\omega$-$\mathbf{k}_\parallel$ domain phase vortices. Second, a perfect mirror substrate that blocks transmission, so incident plane waves and pulses only reflect specularly or diffractively. Third, vertical mirror symmetry perpendicular to the periodicity. Having these three prerequisites is sufficient for structures of \emph{any wave system}, whether scalar or vectorial, transverse or longitudinal, electromagnetic or mechanical, to generate STVPs through our approach.

\par For our approach, the incident angle and frequency are chosen where only specular reflection and -1st-order diffraction are supported by the structure [Fig. \ref{fig:1}(b)]. In other words, the incoming waves scatters into only two scattering channels. Resonances supported by the slab in this frequency-momentum region couple inputs through one channel to the other. At the X point of the Brillouin zone, mirror symmetry causes on-resonance scattering to completely transfer energy between channels, enabling \emph{symmetry-protected resonant total-retroreflection}. Specular reflection vanishes at such an $\omega$-$\mathbf{k}_\parallel$ point, but is non-zero elsewhere [Fig. \ref{fig:1}(b)]. This point is hence a singularity analytically predicted to have a surrounding $\pm1$ phase vortex (see Supplementary Materials (SM) \cite{sm} for proof). Illumination by a pulse that has frequency-momentum components spanning this vortex transfers the winding phase to the pulse, yielding an STVP.

\section*{Simulation Results}

\par
\begin{figure}[ht]
  \centering
  \includegraphics[width=0.5\textwidth]{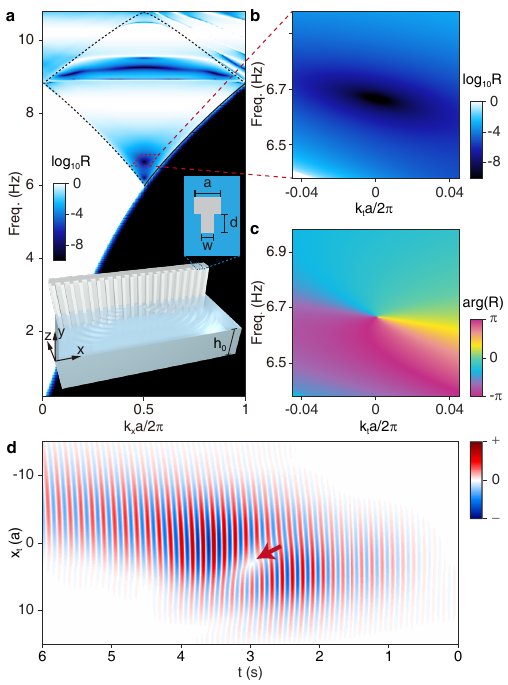}
  \caption{Properties of the resonant mirror-based grating for STVP generation in a liquid-surface-wave system. (a) Simulated momentum-frequency reflectance map by the designed grating. The two-channel region in frequency-momentum space is enclosed by dashed curves, corresponding to the blue-shaded region in Fig. 1(b). The inset provides a schematic of the system and the grating structure. (b) Zoomed-in reflectance map around the singularity  (with zero specular reflectance) at the X point applied to generate the STVP. The frequency of the singularity is about 6.67 Hz. (c) Corresponding reflection phase map showing a $2\pi$ phase vortex around the singularity. (d) Spatiotemporal pulse profile obtained by Fourier transforming the reflectance and phase maps with an assumed incident Gaussian pulse centered at the singularity frequency, showing an STVP. }
  \label{fig:2}
\end{figure}

 Compared to past studies, our approach eliminates the need for complex setups like spatial light modulators or algorithm-optimized structures. To demonstrate universality, we apply it to a liquid-surface-wave system \cite{dingemans1997water, hu2005refraction, zhao2021fast, han2022water, qin2023superscattering, zeng2023achieving}. The applied structure is a hard-boundary-based rectangular grating periodic in the $x$ direction [Fig. \ref{fig:2}(a) inset], leading to well-defined in-plane wavevector $k_x$. The grating's rectangular shape provides the necessary mirror symmetry with respect to planes parallel to the $y$-$z$ plane, and it enables diffraction.


\par Free surface wave propagation follows the dispersion relation \cite{dingemans1997water}:
{\par\nobreak\noindent}
\begin{equation*}\label{eqn:wave_dispersion}
  \omega ^{2}=(gk+\sigma k^{3}/\rho )\tanh (kh_{0}).
\end{equation*}
Here, $\omega$ is the angular frequency of waves, $k$ is the free-space wavevector, $g$ is the gravitational acceleration, $\sigma$ is surface tension, $\rho$ is liquid density, and $h_0$ denotes the undeformed liquid depth. This dispersion relation allows determining allowed propagation modes (the cone of radiation continuum), analogous to light cones in photonics. We consider small amplitude (wave amplitude $\ll$ wavelength), low frequency (6-10 Hz) waves where nonlinear and damping effects are minimal.

\par Introducing the hard-boundary corrugated grating [Fig. \ref{fig:2}(a) inset] produces confined resonant guided modes and associated band structures across momentum space \cite{johnson1999guided, fan2002analysis, fan2003temporal, han2022water}. Meanwhile, the grating also folds the cone of radiation continuum, opening diffraction channels in specific $\omega$-$k_x$ regions. Resonance bands naturally overlap with the region with only specular reflection and -1st order diffraction. Within this overlap region, resonances route a portion of the power in the incident plane waves into the diffraction channel instead of the specular reflection one, causing dips in the specular reflectance. At high-symmetry X point ($k_x = \pi/a$, $a$: grating period) and inside this region, total-retroreflection points induced by resonances (where all the incident power is resonantly diffracted and specular reflectance becomes zero) can be found - which act as $\omega$-$k_x$ domain singularities we implement to generate STVPs.

\par Fig. \ref{fig:2}(a) shows the $\omega$-$k_x$-space specular reflectance map for a grating with parameters: $a=20~$mm; grating width $w=0.5a$; and depth $d=1.12a$. The aforementioned two-channel region is enclosed by dashed curves, corresponding to the blue-shaded region in Fig. \ref{fig:1}(b). Within this region, specular reflectance dips indicate resonances. As discussed, there are several frequencies at X point where the specular reflectance approaches zero, corresponding to total retroreflection points. The specular reflectance map has been logarithmically scaled to highlight these points. Fig. \ref{fig:2}(b) zooms in on a total retroreflection point at $f_0=6.67$ Hz. This is the singularity we will utilize for STVP generation. Accordingly, Fig. \ref{fig:2}(c) reveals an enclosing $2\pi$ phase vortex, as predicted. Note that in Figs. \ref{fig:2}(b)-(c), the horizontal axis is $k_{t}$ rather than the in-plane wavevector $k_x$. $k_{t}$ is the transverse component of the reflected wavevector with respect to a center propagation wavevector $\mathbf{k}(\omega_0)$, obtained by the transform $k_{t} = k\sin \{\arcsin (k_{x}/k)-\arcsin [\pi /a/k(\omega_{0})]\}$ where $\omega_0$ is the singularity's angular frequency.

\par The specularly reflected pulse profile modulated by the grating can be predicted by Fourier transforming the reflectance and phase maps in Figs. \ref{fig:2}(a)-(c), assuming an incident Gaussian pulse centered at the singularity frequency $\omega_0$. (Details of the Fourier transform are provided in SM \cite{sm}). Fig. \ref{fig:2}(d) shows the resulting predicted STVP, with a fork-shaped pattern (red arrow) signifying the presence of a 2$\pi$ vortex in the space-time domain, corresponding to the phase vortex observed in Fig. \ref{fig:2}(c). The STVP vortex stems from the phase singularity at the total retroreflection point in the $\omega$-$k_x$ maps.

\section*{Experimental Demonstration and Discussion}

\begin{figure*}[th]%
  \centering
  \includegraphics[width=1\textwidth]{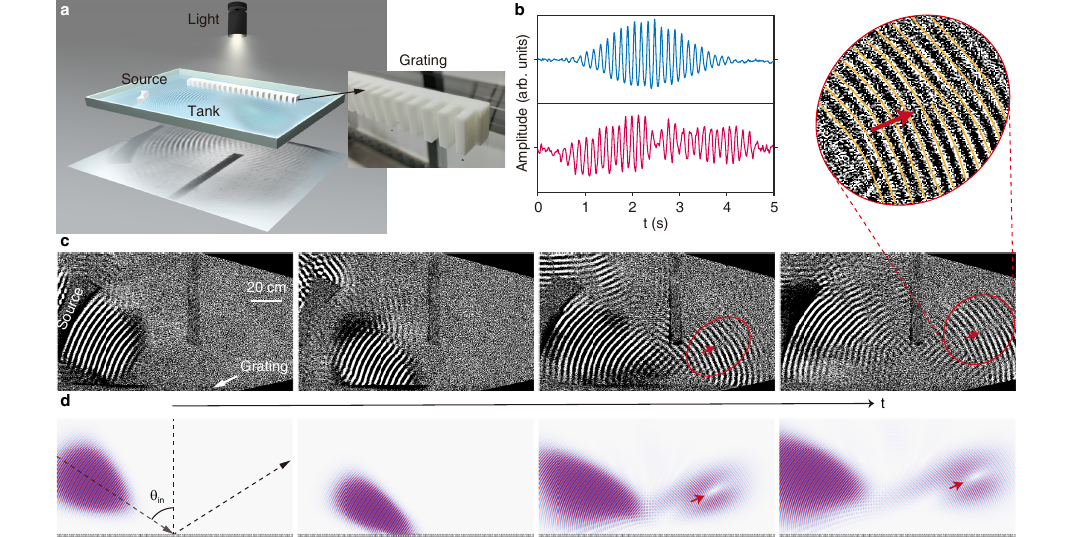}
  \caption{Experimental demonstration of STVP generation using a resonant mirror-based grating. (a) Schematic of the experimental setup. The grating is placed in a transparent tank containing \ce{C2H3ClF2} (1-chloro-1,1-difluoroethane) liquid with a depth of $h_0$. A concave-shaped source is driven by a computer-controlled speaker to emit the incident Gaussian pulse. Inset: photograph of the 3D-printed grating. (b) Measured time signals of the incident Gaussian pulse and reflected STVP, and Fourier-transformed spectra showing normalized spectral intensity of pulses. The signals are sampled at points where the centers of pulses pass. (c) Sequence of time-resolved images showing the reflection process and propagation of the reflected STVP. A retro-reflecting pulse can be observed on the left, while the STVP is in the specular reflecting direction on the right. (d) Simulated time sequence of images for comparison. The central angle of incident pulse, $\theta_{in}$, is about $57.3^{\circ}$.}
  \label{fig:3}
\end{figure*}

We experimentally observed STVP generation using a projected wave tank system, shown in Fig. \ref{fig:3}(a). The system has a transparent tank filled with 1-chloro-1,1-difluoroethane (\ce{C2H3ClF2}) liquid to a depth of $h_0$. A light source above the tank projects wave pattern onto a plane beneath, which can be directly observed or captured on video for analysis. The 3D-printed grating we designed is placed inside the tank [see photograph in Fig. \ref{fig:3}(a)]. A computer-controlled concave speaker source emits Gaussian pulses, with the curvature intentionally designed to shape the wavefronts into spatially Gaussian beams at the grating plane. (An image of a monochromatic Gaussian beam generated by this source at the singularity frequency $\omega_0$ is provided in SM \cite{sm}.) The time signals of the incident Gaussian pulses are sampled at points where the pulse centers pass, as shown by the blue curve in Fig. \ref{fig:3}(b).

\par Fig. \ref{fig:3}(c) shows the time-resolved image sequence of the reflection process and STVP propagation. In the first image (left), a Gaussian pulse is emitted and propagates toward the grating (at the bottom). The incident pulse then impinges on the grating (second image). The third and fourth images show the reflected pulse, and we see that the STVP has already been generated and propagates on the specular side, marked by a red ellipse. A fork-shaped pattern (marked by red arrows), similar to that in Fig. \ref{fig:2}(d), is highlighted by yellow lines. Note that the waves on the left in the last two images are diffraction waves. Supplementary Videos show the complete process \cite{sm}. For reference, Fig. \ref{fig:3}(d) presents the simulated time sequence. Simulations agree well with experiments. See Supplementary Materials for simulation details \cite{sm}. We also extracted the reflected wave time signal at the STVP center [Fig. \ref{fig:3}(b), red curve]. Compared to the incident time signal, an obvious $\pi$ phase jump is visible, corresponding to the frequency domain singularity. Overall, the experiments are consistent with theoretical expectations.


\par
\begin{figure*}[th]%
  \centering
  \includegraphics[width=1\textwidth]{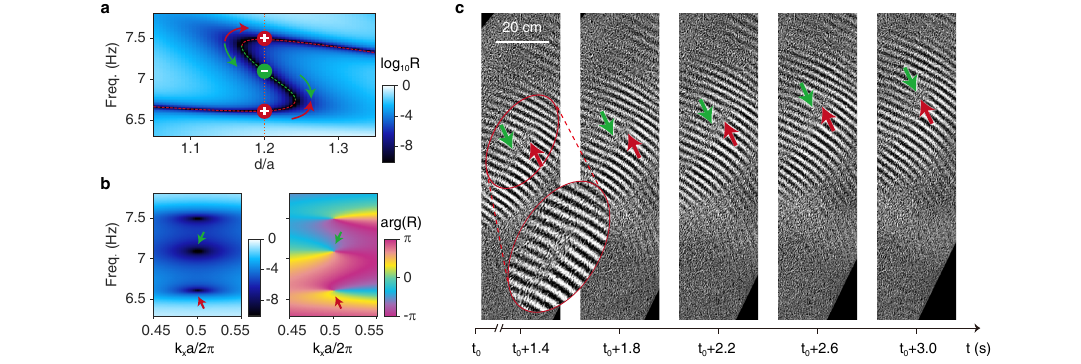}
  \caption{Evolution of phase vortices in the parameter space and coexistence of multiple singularities in one pulse. (a) Simulated reflectance spectrum at the X point as a function of the grating depth $d$. As $d$ increases, a singularity with $+1$ topological charge evolves into a $+1$ charge and a pair of $\pm1$ charges, and then the $+1$ charge annihilates with the $-1$ charge, leaving the other $+1$ charge. The total topological charge is conserved in this process. (b) Momentum-frequency reflectance map for $d=1.2a$ (denoted by orange dashed line in \textbf{a}). The corresponding reflection phase map exhibits two phase vortices with $+1$ charges and one with a $-1$ charge. (c) Experimental images of the reflected pulse. It also shows the real-space annihilation process of the $\pm1$ charge pair which correspond to the $\omega$-$k_x$-domain vortices at $\sim6.7$ Hz and $\sim7.2$ Hz.}
  \label{fig:4}
\end{figure*}

 With single STVP generation demonstrated, we next explore creating multiple interacting singularities within a single spatiotemporal pulse. As emphasized above, our STVP generation approach requires only symmetry, resonance, and diffraction. This enables multiple $\omega$-$k_x$-domain vortices without changing grating symmetry or diffraction orders, and we can generate pulses with multiple spatiotemporal singularities by tuning grating depth and filling fraction. We achieve diverse $\omega$-$k_x$-domain vortex configurations owing to resonance tunability. Fig. \ref{fig:4}(a) shows simulated specular reflectance spectra at the X point as a function of grating depth $d$, with other parameters fixed ($a=30$ mm; $w=0.2a$). As $d$ increases, we observe the complete evolution of total retroreflection points ($\omega$-$k_x$-domain singularities) in the parameter space, including their generation and annihilation. Initially, for $d < 1.16a$, there is a single $+1$ singularity at $\sim6.6$~Hz. At $d \sim 1.16a$, a pair of oppositely charged singularities appear near $7.45$ Hz. With $d$ further increasing, the $-1$ singularity approaches the original $+1$ singularity. At $d \sim 1.24a$, the $-1$ singularity eventually annihilates the original $+1$ singularity at $\sim6.67$~Hz, leaving only the $+1$ singularity at $\sim7.46$~Hz. For $d > 1.24a$, there is again only one $+1$ singularity. The total topological charge is conserved throughout this process.

\par As a concrete example, Fig. \ref{fig:4}(b) depicts the $\omega$-$k_x$-domain specular reflection phase map for $d=1.2a$ [dashed line in Fig. \ref{fig:4}(a)]. Three total-retroreflection points are present, revealing two $+1$-charged phase vortices at $\sim6.7$ Hz and $\sim7.57$ Hz and one $-1$ vortex at $\sim7.2$ Hz. The pairwise vortex generation and coexistence enables observing the real-time evolution of a STVP with multiple singularities.

\par In real space-time, the $\omega$-$k_x$ domain singularities manifest as phase vortices during pulse propagation. Fig. \ref{fig:4}(c) shows the space-time evolution of a generated pulse with multiple singularities for $d=1.2a$. Note that, due to the non-linear dispersion of liquid surface waves, the vortices have different velocities and will dynamically interact. We observe a pair of singularities with opposite charges annihilating during propagation. These annihilating singularities correspond to phase vortices at frequencies of $\sim6.7$ Hz and $\sim7.2$ Hz in Fig. \ref{fig:4}(b) (the lower-frequency two), generated by a Gaussian pulse covering the corresponding $\omega$-$k_x$ region. The time when the center of the incident Gaussian pulse coincides with the grating is denoted as $t_0$. At $t_0+1.4$ s, the two vortices are clearly separated in the pulse, marked by red and green arrows. Their opposing topological charges are evident from the fork-shaped patterns. At $t_0+2.2$ s, they approach each other, still exhibiting fork patterns. However, by $t_0+3.0$ s, the singularities mutually annihilate, indicated by disappearance of the forks. Note that only the specular reflection side field profiles are shown (see Supplementary Videos for full process). The ability to produce complex pulses with multiple singularities highlights the versatility of our general diffractive technique.

\section*{Conclusions}
In summary, we have introduced a general diffractive approach to generate spatiotemporal vortex pulses in diverse wave systems using just resonant planar periodic structures. By harnessing universal wave phenomena of symmetry, diffraction and resonance, phase vortices can be synthesized in the $\omega$-$\mathbf{k}_\parallel$ domain without complex designs. Through the natural reciprocal relationship, this controlled modulation in $\omega$-$\mathbf{k}_\parallel$ domain manifests directly as structured STVP waveforms. We demonstrated the approach with liquid surface waves, enabling on-demand generation and propagation of STVPs. This establishes a versatile framework for wavefront manipulation with potential innovations across disciplines. Further efforts are still needed to fully investigate emerging opportunities, and our approach expands the toolkit for investigating novel spatiotemporal degrees of freedom. Looking forward, we hope this work encourages multi-disciplinary pursuits of fundamental questions and applications related to spatiotemporally-structured pulses \cite{zdagkas2021spatio, wan2022toroidal, zdagkas2022observation, shen2023roadmap}, angular momenta \cite{bliokh2019spin, shi2019observation}, and singularities \cite{ge2021observation, bliokh2021polarization, muelas2022observation} in waves.



\begin{acknowledgments}
\section*{Acknowledgments}
\paragraph*{Funding:}
The authors acknowledge the support of National Key Research and Development Program of China (2022YFA1404800, 2021YFA1400603); National Natural Science Foundation of China (No. 12234007 and No. 12221004); Major Program of National Natural Science Foundation of China (91963212); Science and Technology Commission of Shanghai Municipality (22142200400, 21DZ1101500, 2019SHZDZX01); China Postdoctoral Science Foundation (\grant{2022M720810}, \grant{2022TQ0078}).
This work is also supported by the Research Grants Council of Hong Kong through grants (AoE/P-502/20, A-HKUST601/18) and the Croucher Foundation (CAS20SC01).
\paragraph*{Authors contributions:}
All the authors discussed, interpreted the results and conceived the theoretical framework. W.L. and J.Z. conceived the basic idea of the work. Z.C. extended the model to liquid surface waves and designed the simulations. Z.C., W.L., and L.S. designed the experiments. Z.C. performed the experiments. L.S., C.T.C. and J.Z. supervised the research and the development of the manuscript. Z.C., W.L., L.S., and C.T.C. wrote the draft of the manuscript, and all authors took part in the discussion, revision and approved the final copy of the manuscript. Z.C. and W.L. contributed equally to this work.
\end{acknowledgments}





%

\end{document}